# Support Aggregate Analytic Window Function over Large Data by Spilling

**Xing Shi and Chao Wang**

Guangdong University of Technology, Guangzhou, Guangdong 510006, China
North China University of Technology, Beijing 100144, China

**Abstract.** Analytic function, also called window function, is to query the aggregation of data over a sliding window. For example, a simple query over the online stock platform is to return the average price of a stock of the last three days. These functions are commonly used features in SQL databases. They are supported in most of the commercial databases. With the increasing usage of cloud data infra and machine learning technology, the frequency of queries with analytic window functions rises. Some analytic functions only require const space in memory to store the state, such as SUM, AVG, while others require linear space, such as MIN, MAX. When the window is extremely large, the memory space to store the state may be too large. In this case, we need to spill the state to disk, which is a heavy operation. In this paper, we proposed an algorithm to manipulate the state data in the disk to reduce the disk I/O to make spill available and efficient. We analyse the complexity of the algorithm with different data distribution.

## 1. Introducion

In this paper, we develop novel spill techniques for analytic window function in SQL databases. And discuss different various types of aggregate queries, e.g., COUNT, AVG, SUM, MAX, MIN, etc., over a relational table. Aggregate analytic function, also called aggregate window function, is to query the aggregation of data over a sliding window. It is widely used in the database system and supports a lot of online analytical processing platforms. For example, a trader's daily query is to ask the average price of a stock over the last three days. Most of the commercial databases did a wonderful job on optimising the performance of such queries. Nowadays, the amount of data has increased dramatically and the window of the query is much larger than the common sense we have before. Some analytic functions only require const space in memory to store the state, such as SUM, AVG, while others require linear space, such as MIN, MAX. When the window is extremely large, the memory space to store the state may be too large. In this case, we need to spill the state to disk, which is a heavy operation. In the following sections, we proposed an algorithm to manipulate the state data in the disk to reduce the disk I/O to make spill available and efficient. We analyse the complexity of the algorithm with different data distribution. We will introduce the concept based on Table. 1. In Table. 1, there are three columns and seven rows. We can treat the first (Student ID) or the second column (Name) as key and the Third column (Score) as value.

## 2. Background

### 2.1. *Windows*

An example simple AVG function over a row-based bounded window is Query 1,

SELECT StudentID, Score, AVG(Score) OVER (ROW BETWEEN 2 PRECEDING AND 1 FOLLOWING);

It means that we want to compute the average score of each student's previous two students, one following student and himself in the default order of the current table.

**Table 1.** Student List.

| StudentID | Name | Score |
|---|---|---|
| 000000001 | David | 90 |
| 000000002 | Justin | 70 |
| 000000003 | Alice | 89 |
| 000000004 | Bob | 80 |
| 000000005 | Lucy | 81 |
| 000000006 | Lily | 75 |
| 000000007 | Ray | 86 |

**Table 2.** The result of Query 1.

| StudentID | Score | AVG(Score) |
|---|---|---|
| 000000001 | 90 | (90+70)/2 |
| 000000002 | 70 | (90+70+89)/3 |
| 000000003 | 89 | (90+70+89+80)/4 |
| 000000004 | 80 | (70+89+80+81)/4 |
| 000000005 | 81 | (89+80+81+75)/4 |
| 000000006 | 75 | (80+81+75+86)/4 |
| 000000007 | 86 | (81+75+86)/3 |

The result is Table. 2, take the fifth row as an example, the AVG(Score) we want is the average of the third, forth, fifth, sixth row's Score. We call these rows in the row based window of this query of the current row. There are two types of window, the row based window and the range based window. The row based window is bounded by two rows, the start bound row and the end bound row. The range based window is bounded by the value of the current row and the range given. We take Query 2 as an example,

SELECT StudentID, Score, AVG(Score) OVER (ORDER BY Score RANGE BETWEEN 2 PRECEDING AND 1 FOLLOWING);

The first step is to sort the table rows by Score, the result is Table. 3. The second step is to compute the average based on the value of each row and the given range, as Table. 4.

**Table 3.** Sorted by Score.

| StudentID | Score |
|---|---|
| 000000002 | 70 |
| 000000006 | 75 |
| 000000004 | 80 |
| 000000005 | 81 |
| 000000007 | 86 |
| 000000003 | 89 |
| 000000001 | 90 |

**Table 4.** The result of Query 2.

| StudentID | Score | AVG(Score) |
|---|---|---|
| 000000002 | 70 | 70 |
| 000000006 | 75 | 75 |
| 000000004 | 80 | (80+81)/2 |
| 000000005 | 81 | (80+81)/2 |
| 000000007 | 86 | 86 |
| 000000003 | 89 | (89+90)/2 |
| 000000001 | 90 | (89+90)/2 |

For the sixth row, we can see that the value is 89, and the given range is BETWEEN 2 PRECEDING AND 1 FOLLOWING, which means [89-2, 89+1] -> [87, 90], only the current row and the seventh row are in the range. Thus, the result is (89+90)/2.

Also, we can classify the window based on its start boundary and end boundary,

**Table 5.** Type of windows.

| Window Type | Start Boundary | End Boundary |
|---|---|---|

| Cumulative Window | UNBOUNDED PRECEDING | x PRECEDING/CURRENT ROW/x FOLLOWING |
|---|---|---|
| Unbounded Window | UNBOUNDED PRECEDING | UNBOUNDED FOLLOWING |
| Moving Window | x PRECEDING | y PRECEDING/CURRENT ROW/y FOLLOWING |
| | CURRENT ROW | CURRENT ROW/y PRECEDING |
| | x FOLLOWING | y FOLLOWING |

For the cumulative windows, if we switch the start boundary and the end boundary, we can still take it as a cumulative window by reversing it.

### 2.2. *Aggregation Functions*

There are a lot of commonly used aggregation functions. For most of the aggregate functions, we don't need to recompute the aggregation over all the rows in the current window. We can only care about the rows newly entering the window and the rows leaving the window. With the values of these two rows, we can update the previous result of the row above and get the current result. For example, for a sequence 1, 2, 3, 4, 5. If the moving window contains three items, the first three results are 1, 1+2, 1+2+3. When we are trying to get the fourth result, rather than recomputing 2+3+4, we can update the previous result (1+2+3) with the entering item 4 and the leaving item 1. The new fourth result will be (1+2+3)+4-1. So we only need two operations instead of O(window_width - 1).

Here we discuss them one by one and show which of them are the ones focused in this paper. Here a list

> *ANY_VALUE, ARRAY_AGG, AVG, CORR, COUNT, COUNTIF, COVAR_POP, COVAR_SAMP, MAX, MIN, STDDEV_POP, STDDEV_SAMP, STRING_AGG, SUM, VAR_POP, VAR_SAMP*

For all of them except MIN/MAX, we know that the update method is available for all types of windows. But for MIN/MAX, we know that only unbound windows and cumulative windows can be supported by the updating method. For moving windows, because the new MIN/MAX value cannot be easily determined when the previous MIN/MAX leaves the window, the results cannot be generated by the updating method.

## 3. **A straightforward algorithm for the MIN/MAX over moving windows without spilling**

We know that it doesn't work if we only use constant space in memory to store the necessary state data like SUM/AVG to support MIN/MAX. The brute-force method is to recompute the MIN/MAX over the data in the current window. But it is obviously not a good idea. From this section, we take MAX as an example to discuss. Here we use the fact that the new big number can deprecate the previous small number to reduce the necessary space.

We take a sequence of numbers as an example 7,8,9,6,4,5,3,2,1. If we have a three-item moving window (one previous, current, one following), the result will be 8,9,9,9,6,5,5,3,2.

When an item is entering the window, it is easy to update the max when the new item is larger than the current max. The problem is when an item is leaving the window. There are two situation,

(1) The leaving item is not the current max, which means the leaving item is smaller than the current max, then we need to do nothing. The new max is the current max.
(2) The leaving item is the current max, which means the new max is smaller than the current max, and we need to pick a new max. Here we need to remember the second largest one.

We noticed that if the items we remembered are smaller than the new entering one, all the smaller items can be discarded, only the new entering one and the items larger than it need to be remembered.

Then we have a general algorithm. We use a list to remember the necessary items. Each time when there is a new item entering the window, we remove all the items smaller than the entering item. And then add the entering window at the end of the list. Update the max if the entering item is larger than the current max. Each time when an item is leaving the window, compare the first item of the list with the leaving item. If they are the same, remove the first item and the max will be the newly first item in the list. If they are not, which means the leaving item is smaller than the first item, the max won't change.

We can still use the previous example, 7,8,9,6,4,5,3,2,1. The remembered list, we noted as pair (Entering Item -> Remembered List), will be (7 -> 7), (8 -> 8), (9 -> 9), (6 -> 9,6), (4 -> 9,6,4), (5 -> 9,6,5), (3 -> 9,6,5,3), (2 -> 9,6,5,3,2), (1 -> 9,6,5,3,1).

For the sequence with duplicated items, we can also remember the count of each item to reduce the space. For example, 9,9,9,8,8,8,8,6,5,4,4 can be stored as 9*3+8*4+6+5+4*2.

## 4. **A spilled algorithm**

Because the remembered list's space complexity is O(width), when the window is extremely large, the size of the remembered list can be too large to fit in the memory. In this case, we need to spill the list to the disk. We discuss a way to manipulate the pages on the disk in this section.

When spilling, we cut the list into pages and store them on the disk by the order of their position on the list. It is obvious that the remembered list is a decreasing sequence, so we only need to care about the first page when an item is leaving the window. For the entering item, we need to load the last page first to compare if there is an item larger than the entering item. If not, we need to discard the last page and load the page before the last one, until we find the item larger than the entering one, and add the entering one after it. If no such item is found, the new entry is the only item that needs to be remembered.

Here we have an improvement. If we can store all the pairs (first item, last item) of all the pages in the memory, we can easily find the page that contains the smallest item which is larger than the new entering item, and reduce the disk I/O significantly (from O(window_witdh) to O(1)). If the list of pairs is too large as well, it can be also spilled into the disk, and the pages on the disk have a relation in tree structure. The disk I/O is from O(window_witdh) to O(logwindow_witdh). The space complexity does not change, it is from O(window_witdh) to O(window_witdh+window_witdh/page_size+window_witdh/page_size^2+...)=O(window_witdh)

## 5. **Analysis on Data Distribution**

Assume the given window width is d, the data are equivalent to a list $l_d$ of items with length d, and the remembered list length is x. Then there is a function, $x = f(l_d)$. Also, the expected length of the remembered list is a function of window width d, $E(x)=g(d)$ and k-permutations of n is noted as $P(n,k)$

### 5.1. *Uniform Distribution*

When the items are evenly distributed, a naive method to compute the expected length of the remembered list E(x) is to have all the permutations with length d and compute the average of the corresponding x.

$$E(x) = E(f(l_d)) = SUM(f(l_d))/COUNT(l_d)$$

When d=1, E(x)=1, we note it as g(1)=1. When d=2, E(x)=1.5, thus, g(2)=1.5.

For d=n, we consider the position of the largest number, p_large.

If p_large=n, x=1, the count of l_n meet the condition P(n-1, n-1)P(0,0)

If p_large=n-1, E(x)=2=g(1)+1, the count is P(n-1,n-2)P(1,1)

If p_large=n-2, E(x)=g(2)+1, the count is P(n-1,n-3)P(2,2)

…

If p_large=1, Ex=g(n-1)+1 the count is P(n-1,0)P(n-1,n-1)

Thus, g(n)=(P(n-1,n-1)P(0,0)(g(0)+1)+P(n-1,n-2)P(1,1)(g(1)+1)+P(n-1,n-3)P(2,2)(g(2)+1)+...+ P(n-1,1)P(n-2,n-2)(g(n-2)+1)+P(n-1,0)P(n-1,n-1)(g(n-1)+1))/P(n,n)

Then, $g(n+1)=g(n)+(SUM_{k=0}^{n}(P(n,k))-n*SUM_{k=0}^{n-1}(P(n-1,k)))/(n+1)!$

### 5.2. *Normal Distribution and Worst Cases*

For the normal distribution, we can have a conclusion by the analysis above in section 5.1 that if there is a mapping from the normal distribution to the uniform distribution and keep the ordering of the items, the analysis result keeps the same. We can also have the same conclusion for other distributions.

The worst case is that the input is a decreasing sequence. In this case, x is the same as d. On the other hand, the best case is that the input is an increasing sequence, and x is always 1.

## 6. **Related Work**

The research of [2, 8] focuses on the evaluation optimization of analytic window functions. They propose FS, HS and SS used for tuple reordering. To minimise the total number of FS operations needed, [2] uses the properties of WPKs and WOKs, and clusters window functions into Ordering Groups, while [8] uses cover sets. [2] also worked on predicate pushdown for ranking functions and parallel execution of a single window function, which are also optimizations. [1, 3, 4] worked on the GROUP BY extensions of the optimization techniques, each tuple group collapsing into a single tuple. However, these cannot be used to evaluate the analytic function directly. GROUP BY clause can be extended to GROUPING SETS, ROLLUP and CUBE operations. [2, 6, 7] provided optimization frameworks for the intermediate results held by infer grouping properties and the ordering.